\documentclass[sigconf,screen]{acmart}

\usepackage{booktabs}
\usepackage{multirow}
\usepackage{makecell}
\usepackage{subfigure}
\usepackage[misc]{ifsym} 
\usepackage{balance}
\usepackage{listings}
\usepackage{color, xcolor}
\usepackage{algorithm2e}
\usepackage{diagbox}
\usepackage{tcolorbox}
\usepackage{stfloats}
\AtBeginDocument{%
  \providecommand\BibTeX{{%
    \normalfont B\kern-0.5em{\scshape i\kern-0.25em b}\kern-0.8em\TeX}}}

\setcopyright{acmcopyright}
\acmYear{2025}
\copyrightyear{2025}
\setcopyright{acmlicensed}
\acmConference[FSE '25]{33rd ACM International Conference on the Foundations of Software Engineering}{June 23--28, 2025}{Trondheim, Norway}
\acmBooktitle{33rd ACM International Conference on the Foundations of Software Engineering (FSE '25), June 23--28, 2025, Trondheim, Norway}
\acmDOI{10.1145/3696630.3728492}
\acmISBN{979-8-4007-1276-0/25/06}

\begin{document}
\begin{sloppypar}
	
	\author{Lingzhe Zhang$^{\dag}$}
	\affiliation{%
		\institution{Peking University}
		\city{Beijing}
		\country{China}}
	\orcid{0009-0005-9500-4489}
	\email{zhang.lingzhe@stu.pku.edu.cn}
	
	\author{Yunpeng Zhai$^{\dag}$}
	\thanks{$^{\dag}$Equal contribution}
	\affiliation{%
		\institution{Alibaba Group}
		\city{Beijing}
		\country{China}}
	\orcid{0000-0002-3344-4543}
	\email{zhaiyunpeng.zyp@alibaba-inc.com}
	
	\author{Tong Jia$^{\ast}$}
	\thanks{$^{\ast}$Corresponding author}
	\affiliation{%
		\institution{Peking University}
		\city{Beijing}
		\country{China}}
	\orcid{0000-0002-5946-9829}
	\email{jia.tong@pku.edu.cn}
	
	\author{Xiaosong Huang}
	\affiliation{%
		\institution{Peking University}
		\city{Beijing}
		\country{China}}
	\orcid{0009-0003-3462-5324}
	\email{hxs@stu.pku.edu.cn}
	
	\author{Chiming Duan}
	\affiliation{%
		\institution{Peking University}
		\city{Beijing}
		\country{China}}
	\orcid{0009-0008-4422-6323}
	\email{duanchiming@stu.pku.edu.cn}
	
	\author{Ying Li$^{\ast}$}
	\affiliation{%
		\institution{Peking University}
		\city{Beijing}
		\country{China}}
	\orcid{0000-0002-6278-2357}
	\email{li.ying@pku.edu.cn}
	
	\renewcommand{\shortauthors}{Lingzhe Zhang et al.}

\title[AgentFM: Role-Aware Failure Management for Distributed Databases with LLM-Driven Multi-Agents]{AgentFM: Role-Aware Failure Management for \\ Distributed Databases with LLM-Driven Multi-Agents}

\begin{abstract}
	Distributed databases are critical infrastructures for today’s large-scale software systems, making effective failure management essential to ensure software availability. However, existing approaches often overlook the role distinctions within distributed databases and rely on small-scale models with limited generalization capabilities. In this paper, we conduct a preliminary empirical study to emphasize the unique significance of different roles. Building on this insight, we propose \textbf{AgentFM}, a role-aware failure management framework for distributed databases powered by LLM-driven multi-agents. AgentFM addresses failure management by considering system roles, data roles, and task roles, with a meta-agent orchestrating these components. Preliminary evaluations using Apache IoTDB demonstrate the effectiveness of AgentFM and open new directions for further research.
\end{abstract}



\maketitle

\section{Introduction}

The distributed databases, such as Google Spanner~\cite{corbett2013spanner}, Alibaba OceanBase~\cite{yang2022oceanbase}, TiDB~\cite{huang2020tidb}, and Apache IoTDB~\cite{wang2020apache}, have become integral components of cloud infrastructures, handling vast volumes of data~\cite{kang2022separation, zhang2024time}.

However, these systems frequently encounter anomalies such as system failures and performance degradation, leading to significant financial losses. For example, Alibaba Cloud faces Intermittent Slow Queries (iSQs)~\cite{ma2020diagnosing}, leading to billions of dollars in annual losses. Amazon reports that even a 0.1-second delay in loading caused by database anomalies can lead to a 1\% increase in financial losses~\cite{zhang2024multivariate}. Therefore, it is crucial to detect system failures in real time, analyze the root causes of these failures, and automatically remediate them.

System traces, metrics, and logs capture the states and critical events of active processes, making them essential for managing software failures. These data sources provide insights into both normal operations and deviations signaling potential failures. Leveraging their multimodal nature, recent research has enhanced anomaly detection and diagnosis in complex systems~\cite{zhang2023robust, lee2023eadro, zhang2024survey, zhang2024reducing, zhao2021identifying, lee2023heterogeneous, zhang2022deeptralog, huang2023twin, zheng2024multi, lin2024root, zhang2024towards, zhang2025xraglog, zhang2025scalalog}. For example, Eadro~\cite{lee2023eadro} integrates anomaly detection and root cause localization using multi-source data, while AnoFusion~\cite{zhang2023robust} employs unsupervised multimodal failure detection in microservices. MSTGAD~\cite{huang2023twin} combines all three data types with attentive multimodal learning for graph-based anomaly detection. Studies also explore two-source combinations, such as DeepTraLog~\cite{zhang2022deeptralog}, which pairs traces with logs using a GGNN-based model, and SCWarn~\cite{zhao2021identifying}, which identifies anomalies via heterogeneous metrics and logs. Although the effectiveness of these methods has been demonstrated in specific scenarios, they face several practical challenges when applied to distributed databases:

\begin{itemize}
	\item \textbf{Role Awareness.} In distributed databases, nodes play different roles, each with varying levels of importance. Effective failure management often requires coordination and collaboration among multiple nodes. However, current methods completely overlook these roles, leading to suboptimal fault detection, inaccurate root cause diagnosis, and an inability to provide critical repair strategies.
	\item \textbf{Application Limitations.} Existing failure management models lack generalizability. For instance, these models are typically trained on specific systems and encounter concept drift when applied to new systems. Moreover, since these methods often frame the task as a classification problem, the resulting outputs lack interpretability, which is crucial for Operations and Control Engineers (OCEs) to effectively resolve issues.
\end{itemize}

To address the first challenge, we conduct a detailed analysis of the various roles within a distributed database and the entire failure management process. We ultimately identify three distinct role categories: system roles, data roles, and task roles. System roles represent the various roles inherent in the distributed database itself (e.g., leader nodes, follower nodes). Data roles refer to the different data sources involved in failure management, while task roles represent the various tasks that need to be executed during failure management.

To tackle the second challenge, we adopt a large language model (LLM)-based approach. While many existing LLM-based failure management solutions have been proposed, a significant number of them do not utilize the three types of data sources mentioned above~\cite{zhang2024automated, goel2024x, zhang2024lm, ahmed2023recommending, roy2024exploring}. Moreover, some approaches that do incorporate these data sources fail to account for the role-based structure within distributed databases~\cite{zhang2024mabc, shetty2024building, hrusto2024autonomous}. Therefore, we propose a role-aware, LLM-driven multi-agent approach that integrates the characteristics of these roles and the three data sources.

Building on these insights, we introduce AgentFM, a comprehensive role-aware failure management framework for distributed databases, powered by LLM-driven multi-agent systems. AgentFM integrates the unique characteristics of distributed database roles with the rich multimodal data sources typically encountered in failure management, such as system traces, metrics, and logs. By employing a multi-agent architecture, AgentFM facilitates specialized agents for each role—system roles, data roles, and task roles—ensuring a more nuanced and effective approach to failure detection, diagnosis, and resolution.

We conduct preliminary experiments on Apache IoTDB~\cite{wang2020apache}, a distributed time-series database system, to assess the effectiveness of AgentFM in failure detection and root cause analysis. Furthermore, we manually verified the accuracy and validity of the generated mitigation solution.

\textit{Contributions.} The contributions of this paper are threefold. First, we conduct a preliminary empirical study highlighting the varying significance of different roles in failure management. Second, we propose AgentFM, a role-aware failure management framework for distributed databases with LLM-driven multi-agents. Third, we provide a preliminary evaluation of AgentFM on Apache IoTDB, demonstrating its feasibility and effectiveness.

\begin{figure*}[htb]
	\begin{minipage}[b]{1.0\textwidth}
		\centering
		\includegraphics[width=\textwidth]{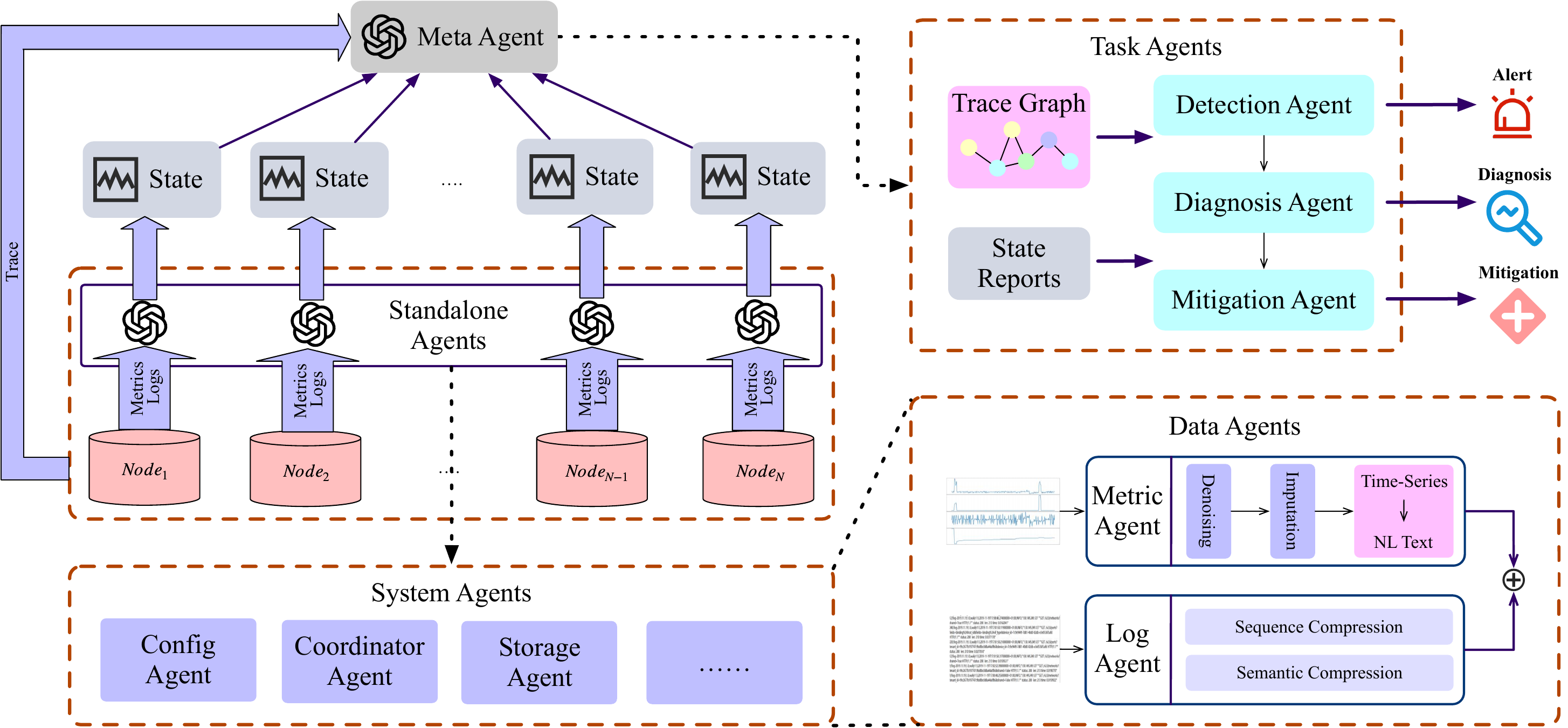}
		\caption{AgentFM Architecture}
		\label{fig: architecture}
	\end{minipage}
\end{figure*}

\section{Preliminary Empircal Study}

In this section, we conduct a preliminary empirical study using Apache IoTDB~\cite{wang2020apache}, focusing on the significance of different roles in the failure management process.

\subsection{System Roles}

We manually injected anomalies (excessive data export) into Apache IoTDB during runtime and evaluate the anomaly detection performance (precison, recall and f1-score) for each node using PLELog~\cite{yang2021semi}, which is a state-of-the-art log-based anomaly detection method.

\begin{table}[h]
	\centering
	\caption{Anomaly Detection Results for Each Node}
	\label{tab: detecting-single-leader-export}
	\begin{tabular}{ccccccc}
		\toprule
		& $Node_{1}$ & $Node_{2}$ & $Node_{3}$ & $Node_{4}$ &  $Node_{5}$ &  $Node_{6}$ \\
		\midrule
		\textbf{p} & 39.68\% & 34.35\% & 59.64\% & 31.17\% & 77.60\% & 89.42\% \\
		\textbf{r} & 99.01\% & 100.00\% & 98.02\% & 100.00\% & 96.04\% & 92.08\% \\
		\textbf{f1} & 56.66\% & 51.14\% & 74.16\% & 47.53\% & 85.84\% & 90.73\% \\
		\bottomrule
	\end{tabular}
\end{table}

As shown in Table~\ref{tab: detecting-single-leader-export}, $Node_{6}$ achieves the best anomaly detection performance due to hosting the largest number of leader partitions. In contrast, the detection performance on other nodes is relatively suboptimal. This observation underscores the differing significance of various system roles.

\subsection{Data Roles}

We further conduct anomaly diagnosis classification experiments on Apache IoTDB using both metrics data and log data. To ensure fairness, we do not adopt state-of-the-art methods; instead, we implement a simple classification algorithm based on the Dynamic Time Warping (DTW) algorithm.

\begin{table}[htbp]
	\centering
	\caption{Anomalies that can be Classified by Metrics and Logs}
	\label{tab: anomalies-can-ba-classified}
	\begin{tabular}{c|c|c|c|c|c}
		\toprule
		\textbf{Source} & CPU & Memory & Export & Import & Configutation \\
		\midrule
		Metrics & $\checkmark$ & $\checkmark$ & & & \\
		\midrule
		Logs & & & $\checkmark$ & $\checkmark$ & $\checkmark$ \\
		\bottomrule
	\end{tabular}
\end{table}

As shown in Table~\ref{tab: anomalies-can-ba-classified}, anomalies with an F1-score above 50\% are considered identifiable by the corresponding data type. The results reveal that metrics are particularly effective at detecting resource anomalies, such as CPU and memory saturation, while logs excel at identifying internal database issues, including excessive data export/import and configuration errors. This observation highlights the distinct significance of different data roles.

\section{Methodology}

Our preliminary empirical study highlights that the effectiveness of failure management varies significantly across different roles in distributed databases. Consequently, it becomes crucial to assign varying levels of importance to different roles at different times and adopt role-specific operations to optimize outcomes.

In this section, we introduce \textbf{AgentFM}, a role-aware failure management framework for distributed databases powered by LLM-driven multi-agents. Figure~\ref{fig: architecture} illustrates the architecture of AgentFM, which comprises three types of agents corresponding to distinct roles: system agents, data agents, and task agents. System agents represent the various roles of nodes within the distributed database, data agents handle information from diverse data sources, and task agents focus on executing specific failure management tasks. Notably, since system agents and data agents operate independently on each node, they are collectively referred to as standalone agents. Additionally, the framework incorporates a meta-agent, which is responsible for orchestrating, adapting, aggregating, and ultimately producing the final results of the agents' operations, ensuring cohesive and efficient failure management.

The framework operates by first extracting critical information from different system agents using the corresponding data agents. This information is then transmitted to a centralized meta-agent for unified decision-making. The decision-making process employs trace data as a cohesive framework to integrate key insights across nodes. Finally, the appropriate task agents execute the required failure management operations, completing the process efficiently and effectively.

\subsection{System Agents}

Different distributed databases assign dynamic system roles that may change during runtime (e.g., leader re-election after node failure). To handle this, we designed an adaptive mechanism for system agents within the Meta Agent.

\begin{figure}[h]
	\begin{minipage}[b]{1.0\linewidth}
		\centering
		\includegraphics[width=\linewidth]{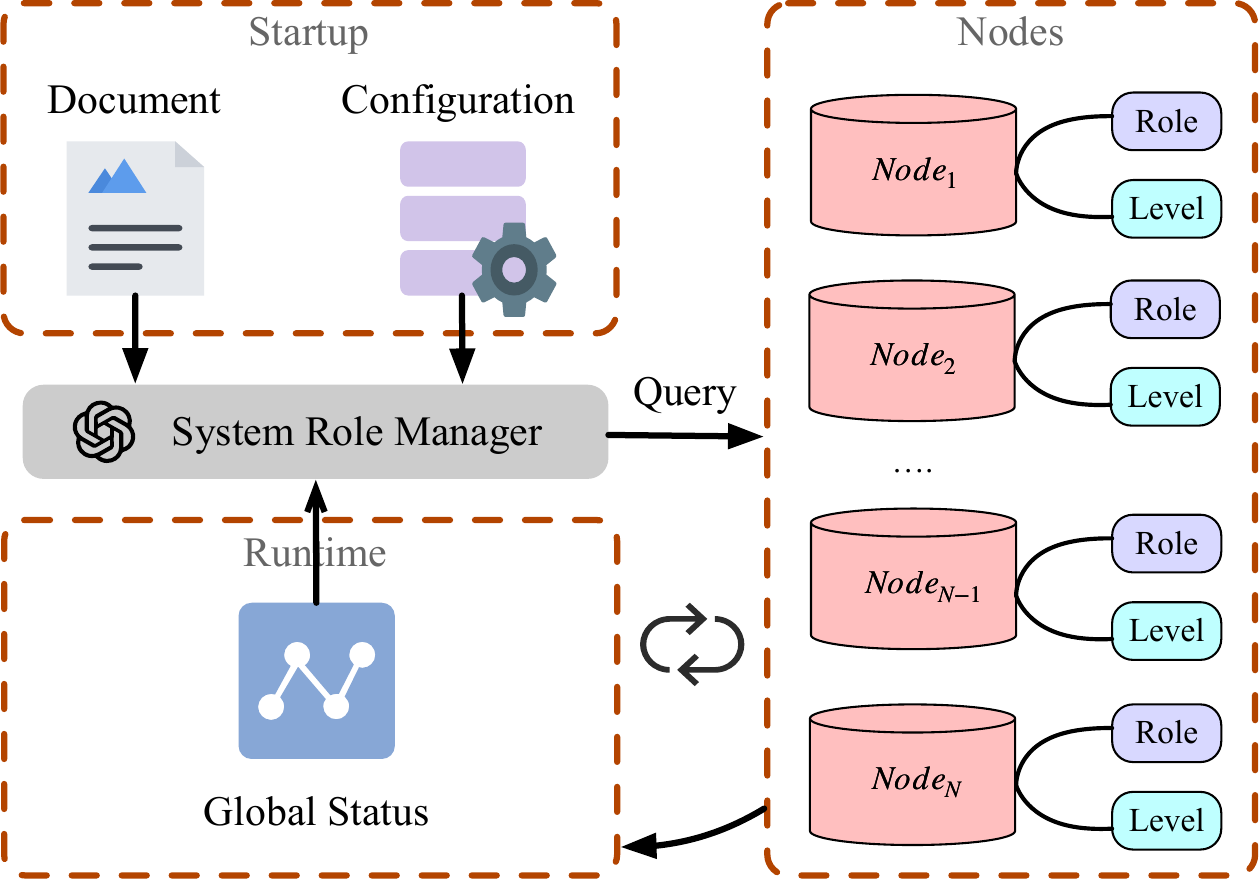}
		\caption{System Agents Adaptation Workflow}
		\label{fig: system-agent}
	\end{minipage}
\end{figure}

As illustrated in Figure~\ref{fig: system-agent}, the core component is the System Role Manager, which initializes by analyzing configuration files and system docs to identify each node's role and importance. At runtime, it periodically queries the system state and updates roles accordingly.

Based on this process, each node instantiates a corresponding system agent—such as a Config Agent, Coordinator Agent, or Storage Agent—each embedding multiple data agents to handle specialized tasks.

\subsection{Data Agents}

In this paper, we define two types of data agents: the Metric Agent and the Log Agent. Each adopts a distinct approach to extract key information from the raw metrics data and log data, respectively.

\textbf{Metric Agent: } The overall workflow of the Metric Agent begins with simple data preprocessing, which includes noise removal and imputation of missing values. After preprocessing, the multivariate time-series data is converted into natural language descriptions using a large model. The generated natural language descriptions capture key aspects of the data, including time intervals, fluctuation trends, and anomaly points. 

Formally, let $\mathbf{M} = \{m_1, m_2, ..., m_n\}$ represent the raw multi-dimensional metrics data, where each $m_i$ corresponds to a time-series for a specific metric. The preprocessing step can be represented as Equation~\ref{eq: metric-preprocess}, where $\mathbf{M}_{\text{p}}$ is the denoised and imputed version of $M$.

\begin{equation}
	\mathbf{M}_{\text{p}} = Preprocess(\mathbf{M}) = \{m'_1, m'_2, ..., m'_n\}
	\label{eq: metric-preprocess}
\end{equation}

Next, the processed data is converted into natural language descriptions using a large language model $\mathcal{L}$, as shown in Equation~\ref{eq: metric-nlp}, where $\mathbf{D}_{\text{nl}}$ represents the resulting natural language description, which includes information on time intervals, trends, and identified anomalies in the data.

\begin{equation}
	\mathbf{D}_{\text{nl}} = \mathcal{L}(\mathbf{M}_{\text{p}})
	\label{eq: metric-nlp}
\end{equation}

\textbf{Log Agent:} Unlike metrics data, log data is written by developers to monitor the internal state of the system, inherently containing rich semantic information. However, log data often includes substantial redundant information, such as repetitive log entries and messages unrelated to system anomalies.

To address this, the Log Agent comprises two main components: sequence compression and semantic compression. Sequence compression focuses on reducing the raw log sequence by consolidating repetitive log patterns, while semantic compression extracts key operational information from the logs.

\textit{Sequence Compression:} This component utilizes a log-parsing algorithm to transform each log entry into a distinct event template consisting of a static constant part and variable parameters. It then merges consecutive occurrences of identical event templates. Formally, for a given raw log sequence \( L = \{l_1, l_2, \dots, l_N\} \), where \( l_n \) represents an individual log entry, the parsing process transforms \( L \) into \( L_e = \{e_1, e_2, \dots, e_N\} \), where \( e_i \) represents an event template. Consecutive identical events are merged, yielding \( G' = \{e'_1, e'_2, \dots, e'_C\} \), where \( C \ll N \) and \( e'_i = e_i \times c \) represents the event \( e_i \) with a count \( c \).

\textit{Semantic Compression:} This component employs a prompt-based method to convert lengthy log sequences into descriptions of the database's ongoing operations. Formally, given a sequence of raw log entries \( L = \{l_1, l_2, \dots, l_N\} \), the LLM-based summarization transforms and compresses \( L \) into a smaller set of key operational elements \( O = \{o_1, o_2, \dots, o_M\} \), where \( M \ll N \).

\subsection{Task Agents}

After system agents collect necessary data via data agents, the meta agent coordinates task agents to carry out failure management using trace data. This process involves three agent types—detection, diagnosis, and mitigation—which operate sequentially. Detection agents identify anomalies in each time window; upon detection, diagnosis agents locate and classify the issue; finally, mitigation agents propose solutions based on the diagnosis.

Though prompts vary by task, all agents follow a common RAG+CoT approach, using historical data as labeled examples—normal/abnormal for detection, failure types for diagnosis—to guide reasoning during querying.

\section{Preliminary Evaluation}

\subsection{Design}

To evaluate AgentFM, we assess its feasibility and effectiveness in Apache IoTDB. We manually injected 10 types of anomalies, including CPU saturation, IO saturation, memory saturation, network delay increase, network bandwidth limitation, network partition occurrence, workload spikes, accompanying slow queries, excessive data export, and excessive data import. Each anomaly type is injected 20 times.

The evaluation is conducted based on Qwen2.5-72b to assess the results of anomaly detection and diagnosis. The performance is measured using precision, recall, and F1-score metrics. Additionally, the mitigation solutions generated are manually analyzed to determine their usefulness.

\subsection{Results}

We utilize the function call feature to guide the model in generating "yes" or "no" responses to evaluate anomaly detection performance. For failure diagnosis, we manually review the LLM-generated results to identify the classification most closely matching the pre-injected anomalies and assessed the outcomes accordingly.

\begin{table}[h]
	\centering
	\caption{Anomaly Detection \& Diagnosis Results}
	\label{tab: detection-diagnosis-result}
	\begin{tabular}{cccc}
		\toprule
		\textbf{Task} & Precision & Recall &  F1-Score \\
		\midrule
		Anomaly Detection & 95.14\% & 97.03\% & 95.76\%  \\
		Failure Diagnosis & 89.61\% & 87.04\% & 87.62\% \\
		\midrule
	\end{tabular}
\end{table}

As shown in Table~\ref{tab: detection-diagnosis-result}, AgentFM demonstrates promising performance in our preliminary experiments. However, based on our observations, the model struggles to identify correct results when dealing with complex internal anomalies (e.g., excessive data import). We plan to focus on optimizing performance for such scenarios in future work.

\begin{figure}[htbp]
	\centering
	\begin{tcolorbox}[colback=gray!10,
		colframe=black,
		width=\linewidth,
		arc=1mm, auto outer arc,
		boxrule=0.5pt,
		top=2pt, 
		bottom=2pt, 
		left=2pt,
		right=2pt
		]
		\textbf{Increase CPU Resources:}
		
		Iotdb0 frequently experiences high CPU usage, consider increasing the number of CPU cores allocated to the container or raising the CPU limit.
		
		\textbf{Load Balancing:}
		
		If iotdb0 is part of a cluster, consider distributing some of the load to other nodes to achieve load balancing. 
	\end{tcolorbox}
	\caption{Sample Mitigation Solutions from AgentFM}
	\label{fig: mitigation-solution}
\end{figure}

We also conduct a manual analysis of the mitigation solutions recommended by AgentFM. One example is illustrated in Figure~\ref{fig: mitigation-solution}. We extract the key information from the solution and find that these recommendations provide targeted and effective assistance in system remediation.

\section{Conclusion}

This paper investigates the varying significance of different roles in failure management. Building on this insight, we propose AgentFM, a role-aware failure management framework for distributed databases powered by LLM-driven multi-agents. Our preliminary experiments demonstrate the feasibility of AgentFM. In the future, we plan to further enhance AgentFM's capabilities in automating remediation execution.

\section*{Acknowledgement}

This work is supported by Key-Area Research and Development Program of Guangdong Province, China (NO.2020B010164003).

\bibliographystyle{ACM-Reference-Format}
\balance
\bibliography{sample-base}

\clearpage

\end{sloppypar}
\end{document}